\documentclass[12pt]{article}

\usepackage{amssymb}
\usepackage[mathscr]{euscript}
\usepackage{graphicx}
\usepackage[all]{xypic}

\newtheorem{prop}{Proposition}[section]
\newtheorem{rema}{Remark}[section]

\newtheorem{lemm}{Lemma}[section]
\newtheorem{theo}{Theorem}[section]
\newtheorem{coro}{Corollary}[section]
\newtheorem{exem}{Example}[section]

\newcommand{\bbox}{\normalsize {}%
        \nolinebreak \hfill $\blacksquare$ \medbreak \par}

\newcommand{\x}{\mathfrak{s}}

\newcommand{\A}{\Bbb{A}}
\newcommand{\AI}{\Bbb{A}\!\Bbb{I}}
\newcommand{\R}{\Bbb{R}}

\newcommand{\Z}{\Bbb{Z}}

\newcommand{\N}{\Bbb{N}}
\newcommand{\I}{\Bbb{I}}

\title{A representation formula for maps on supermanifolds}
\author{Fr{\'e}d{\'e}ric {\sc H{\'e}lein}\footnote{helein@math.jussieu.fr
Institut de Math{\'e}matiques de Jussieu, UMR 7586 
 Universit{\'e} Denis Diderot -- Paris 7,
Case 7012, 2 place Jussieu
75251 Paris Cedex 5, France}
}

\begin{document}
\maketitle
\section*{Introduction}
The theory of supermanifolds, first proposed by Salam and Strathdee
\cite{SalamStrathdee} as a geometrical framework for
understanding the supersymmetry, is now well understood
mathematically and can be formulated in roughly two different ways:
either by defining a notion of superdifferential structure with
"supernumbers" which
generalizes the differential structure of $\R^p$ and by gluing
together these local models to build a supermanifold. This is the
approach proposed by Dewitt \cite{Dewitt} and Rogers \cite{Rogers}. Alternatively
one can define supermanifolds  as ringed spaces, i.e. objects on which the
algebra (or the sheaf) of functions is actually a superalgebra (or a
sheaf of superalgebras). This point of view was adopted by Berezin
\cite{Berezin}, Le{\u\i}tes \cite{Leites}, Manin \cite{Manin} and was
recently further developped by Deligne and Morgan
\cite{DeligneMorgan}, Freed \cite{Freed} and Varadarajan
\cite{Vara}. The first approach is influenced by
differential geometry, whereas the second one is inspired by algebraic
geometry. Of course all these points of view are strongly related, but
they may lead to some subtle differences (see Batchelor \cite{batchelor}, Bartocci, Bruzzo and
Hern{\'a}ndez-Ruip{\'e}rez \cite{bbh} and Bahraini \cite{bahraini}).\\

\noindent
The starting point of this paper was to understand some implications of
the theory of supermanifolds according to the second point of view \cite{Berezin,Leites, 
Manin,DeligneMorgan,Freed,Vara}, i.e. the one inspired by algebraic geometry. The basic
question is to understand $\R^{p|q}$, the space
with $p$ ordinary (bosonic) coordinates and $q$ odd (fermionic)
coordinates. There is no direct definition nor picture of such a space
beside the fact that the algebra of functions on $\Bbb{R}^{p|q}$
should be isomorphic to ${\cal C}^\infty(\R^p)[\eta^1,\cdots ,\eta^q]$, i.e. the algebra over
${\cal C}^\infty(\R^p)$ spanned by $q$ generators $\eta^1, \cdots ,\eta^q$ which
satisfy the anticommutation relations $\eta^i\eta^j + \eta^j\eta^i =
0$. Hence ${\cal C}^\infty(\R^p)[\eta^1,\cdots ,\eta^q]$ is isomorphic
to the set of sections of the flat vector bundle over $\R^p$ whose
fiber is the exterior algebra
$\Lambda^*\Bbb{R}^q$. To experiment further $\Bbb{R}^{p|q}$ we define
what should be maps from open subsets of $\Bbb{R}^{p|q}$ to ordinary
manifolds. We adopt the provisional definition of an {\em open subset} of $\Omega$ of $\R^{p|q}$
to be a space on which the algebra of functions is isomorphic to ${\cal
  C}^\infty(|\Omega|) [\eta^1,\cdots ,\eta^q]$, where $|\Omega|$ is an
open subset of $\R^p$. So we choose such an open set $\Omega$ and a
smooth ordinary manifold ${\cal N}$ and analyze what should be maps 
$\phi$ from $\Omega$ to ${\cal N}$. Again there is no direct definition of such an object
except that by the chain rule it should define a ring morphism
$\phi^*$ from the ring ${\cal C}^\infty({\cal N})$ of smooth functions on ${\cal N}$ to
the ring ${\cal C}^\infty(|\Omega|)[\eta^1,\cdots ,\eta^q]$. The
morphism property means that 
\begin{equation}\label{ax+}
\forall \lambda,\mu\in \R, \forall f,g\in {\cal C}^\infty({\cal N}),
\quad \phi^*(\lambda f + \mu g) = \lambda \phi^* f + \mu \phi^*g
\end{equation}
and
\begin{equation}\label{axprod}
\forall f,g\in {\cal C}^\infty({\cal N}),
\quad \phi^*(fg) = (\phi^* f)(\phi^*g).
\end{equation}
We restrict ourself to {\em even} morphisms, which means here that we
impose to $\phi^*f$ to be in the even part ${\cal
  C}^\infty(|\Omega|) [\eta^1,\cdots ,\eta^q]_0$ of ${\cal
  C}^\infty(|\Omega|) [\eta^1,\cdots ,\eta^q]$. 
\\

\noindent
In the first section
we prove our main result (Theorem \ref{letheo}) which shows that, for
any even morphism $\phi^*$, there
exists a smooth map $\varphi$ from $|\Omega|$ to ${\cal N}$ and a
family of vector fields $\left(\Xi_x\right)_{x\in |\Omega|}$ depending
on $x\in |\Omega|$ and tangent to ${\cal N}$ and with coefficients in
the commutative subalgebra $\R[\eta^1,\cdots ,\eta^q]_0$ such that
\begin{equation}\label{main}
\forall f\in {\cal C}^\infty({\cal N}),\quad
\phi^* f = (1\times \varphi)^*\left(e^\Xi f\right).
\end{equation}
One may interpret the term $e^\Xi$ as an analogue with odd variables of the standard
Taylor series representation
\[
g(x) = \sum_{k=0}^\infty {\partial ^kg\over (\partial x)^k}(x_0)
{(x-x_0)^k\over k!} = \left(e^{\sum_{i=1}^n(x^i-x^i_0){\partial \over
    \partial x^i}}g\right) (x_0),
\]
for a function $g$ which is analytic in a neighbourhood of $x_0$. We
also show that the vector field $\Xi$ (which is not unique) can be
build as a combination of commuting vector fieds. Then
the rest of this paper is devoted to the consequences of this
result.\\

\noindent
The second section explores in details the
structure behind relation (\ref{main}). First we exploit the fact that one can assume that the
vector fields which compose $\Xi$ commute, so that one can
integrate them locally. This gives us an alternative description of
morphisms. Eventually this study leads us to a
factorization result for all even morphisms as follows. First let us denote by $\Lambda^{2*}_+\R^q$ the
subspace of all even elements of the exterior algebra $\Lambda^*\R^q$
of positive degree
(i.e. $\Lambda^{2*}_+\R^q \simeq \R^{2^{q-1}-1}$). We
construct an ideal ${\cal I}^q(|\Omega|)$ of the algebra ${\cal
  C}^\infty(|\Omega|\times \Lambda^{2*}_+\R^q)$ in such a way that, if we consider the
quotient algebra ${\cal A}^q(|\Omega|):= {\cal C}^\infty(|\Omega|\times \Lambda^{2*}_+\R^q)/
{\cal I}^q(|\Omega|)$, then there exists a canonical isomorphism $T_\Omega^*: {\cal
  A}^q(|\Omega|) \longrightarrow {\cal C}^\infty(|\Omega|\times \Lambda^{2*}_+\R^q)$.
By following the theory of scheme of Grothendieck we
associate to ${\cal A}^q(|\Omega|)$ its spectrum $\hbox{Spec}{\cal A}^q(|\Omega|)$, a kind of geometric
object embedded in $|\Omega|\times \Lambda^{2*}_+\R^q$. 
Then for any even morphism $\phi^*$ from ${\cal
  C}^\infty({\cal N})$ to ${\cal C}^\infty(|\Omega|)[\eta^1,\cdots
,\eta^q]$, there exists a smooth map $\Phi$ from $|\Omega|\times
\Lambda^{2*}_+\R^q$ to ${\cal N}$, such that
\[
\phi^* = T^*_\Omega \circ \Phi^*_{|\star},
\]
where $\forall f\in {\cal C}^\infty({\cal N})$, $\Phi^*_{|\star}f =
f\circ \Phi$ $\hbox{mod}{\cal I}^q(|\Omega|)$. So by dualizing we can think of the
map $\Phi_{|\star}:\hbox{Spec}{\cal A}^q(|\Omega|) \longrightarrow
{\cal N}$ as the restriction of $\Phi$ to $\hbox{Spec}{\cal A}^q(|\Omega|)$.
Hence we obtain an interpretation of a map on a supermanifold
as a function defined on an (almost) ordinary space. This reminds somehow
the theory developped by Vladimirov and Volovich \cite{Vladi}
who represent a map on a superspace as a function depending on many
auxiliary ordinary variables satisfying a system of so-called
"Cauchy--Riemann type equations". However their description in terms of ordinary
functions satisfying first order equations differs from our point of view.\\

\noindent
The last section is devoted to applications of our results for
understanding the use of supermanifolds by physicists. First we
explain briefly how one can reduced the study of maps between two
supermanifolds to the study of maps from a super manifold to an
ordinary one, by using charts. Second we recall why it is necessary to
incorporate the notion of the functor of point (as illustrated in this
framework in \cite{DeligneMorgan, Freed, FP2}) in the definition of a
map $\phi$ between supermanifolds in terms of ring morphisms. Then we
address the simple question of computing the pull-back image $\phi^*f$
of a map $f$ on an ordinary manifold ${\cal N}$ by a map $\phi$
from an open subset of $\R^{p|q}$ to ${\cal N}$. For instance consider a superfield $\phi = \varphi +
\theta^1\psi_1 + \theta^2\psi_2 +\theta^1\theta^2F$ from $\R^{3|2}$
(with coordinates $(x^1,x^2,t,\theta^1,\theta^2)$) to $\R$ and look at
the Berezin integral 
\[
I:= \int_{\R^{3|2}}d^3xd^2\theta\ \phi^*f,
\]
where $f:\R\longrightarrow \R$ is a smooth function. Such a quantity
arises for instance in the action $\int_{\R^{3|2}}d^3xd^2\theta \left({1\over
  4}\epsilon^{ab}D_a\phi D_b\phi + \phi^*f\right)$ and then $f$ plays
the role of a superpotential. Following Berezin's rules the integral
$I$ is equal to the integral over $\R^3$ of the coefficient of
$\theta^1\theta^2$ in the development of $\phi^*f$, which is actually
\begin{equation}\label{but}
\phi^*f = f\circ \varphi + \theta^1 (f'\circ \varphi)\psi_1 + \theta^2
 (f'\circ \varphi)\psi_2 +  \theta^1\theta^2 [(f'\circ \varphi)F -
 (f''\circ \varphi) \psi_1\psi_2],
\end{equation}
so that $I = \int_{\R^3}d^3x[(f'\circ \varphi)F -
 (f''\circ \varphi) \psi_1\psi_2]$. The development (\ref{but}) is well-known
and can be obtained by several approaches. For instance in
 \cite{FreedDeligne} or in \cite{Freed} one computes the coefficient of
$\theta^1\theta^2$ in the development of $\phi^*f$ by the rule
 $\iota^*\left(-{1\over 2}(D_1D_2 - D_2D_1)\phi^*f\right)$, where
 $D_1$ and $D_2$ are derivatives with respect to $\theta^1$ and
 $\theta^2$ respectively and $\iota$ is the canonical embedding $\R^3
 \hookrightarrow \R^{3|2}$. Here we propose a recipe which, I find, is
 simple, intuitive, but mathematically safe for performing
 this computation (this recipe is of course equivalent to the already existing
 rules !). It consists roughly in the following: we reinterpret the relation
 $\phi = \varphi + \theta^1\psi_1 + \theta^2\psi_2
+\theta^1\theta^2F$ as
\begin{equation}\label{depart}
\phi^* = \varphi^*e^{\theta^1\psi_1 + \theta^2\psi_2
+\theta^1\theta^2F} =
\varphi^*(1+\theta^1\psi_1)(1+\theta^2\psi_2) (1 +
  \theta^1\theta^2F),
\end{equation}
where
\begin{itemize}
\item $\psi_1$, $\psi_2$ and $F$ are first order differential
operators which acts on the right, i.e. for instance $\forall f\in
{\cal C}^\infty(\R)$, $\psi_af = df(\psi_a) = f'\psi_a$ and so
$\varphi^*\psi_af = (f'\circ \varphi)\psi_a$
\item $\psi_1$, $\psi_2$ and $F$ are $\Z_2$-graded in such a way that $\phi^*$
  is even, i.e. since $\theta^1$ and $\theta^2$ are odd, $\psi_1$ and
  $\psi_2$ are odd and $F$ is even
\item all the symbols $\theta^1$, $\theta^2$, $\psi_1$, $\psi_2$ and
  $F$ supercommute.
\end{itemize}
Let us use the supercommutation rules to developp (\ref{depart}), we
obtain: $\forall f\in {\cal C}^\infty(\R)$,
\[
\phi^*f = \varphi^*f + \theta^1 \varphi^*\psi_1f + \theta^2
\varphi^*\psi_2f + \theta^1\theta^2\varphi^*F f -
\theta^1\theta^2\varphi^*\psi_1\psi_2 f.
\]
Then we let the first order differential operators act and this
gives us exactly (\ref{but}).\\

\noindent
All these rules are expounded in details in the third section of this
paper. Their justification is precisely based on the results of the
first section.

\section{Even maps from $\R^{p|q}$ to a manifold ${\cal N}$}
Our first task will be to study even morphisms $\phi^*$ from ${\cal
  C}^\infty({\cal N})$ to ${\cal C}^\infty(|\Omega|)[\eta^1,\cdots
,\eta^q]$, i.e. maps between these two superalgebras which satisfy 
(\ref{ax+}) and (\ref{axprod}). Let us first precise the sense of {\em even}.
If $A = A_0 \oplus A_1$ and $B = B_0 \oplus B_1$ are two
$\Bbb{Z}_2$-graded rings with unity, a ring morphism $\phi: B \longrightarrow A$ is
sayed to be {\em even} is it respects the grading, i.e. $\forall b\in B_\alpha$,
$\phi(b)\in A_\alpha$ for $\alpha = 0,1$. In the case
at hand $B = {\cal C}^\infty({\cal N})$ is purely even, i.e. $B_1 =
\{0\}$, and so $\phi^*$ is even if and only if it maps ${\cal
  C}^\infty({\cal N})$ to ${\cal C}^\infty(|\Omega|)[\eta^1,\cdots
,\eta^q]_0$, the even part of ${\cal C}^\infty(|\Omega|)[\eta^1,\cdots
,\eta^q]$. We then say that $\phi$ is an {\em even} map from $\Omega$
to ${\cal N}$. In the following we shall denote by ${\cal C}^\infty(|\Omega|)[\eta^1,\cdots
,\eta^q]$ and ${\cal C}^\infty(|\Omega|)[\eta^1,\cdots ,\eta^q]_0$ respectively by
${\cal C}^\infty(\Omega)$ and ${\cal C}^\infty(\Omega)_0$ and we shall denote by
$\hbox{Mor}({\cal C}^\infty({\cal N}),{\cal C}^\infty(\Omega)_0)$ the set of
even morphisms from ${\cal C}^\infty({\cal N})$ to ${\cal C}^\infty(\Omega)$.\\

\noindent We observe that because of the hypothesis (\ref{ax+}) any such
morphism is given by a finite family $\left(
  a_{i_1\cdots i_{2k}}\right)$ of linear functionals on ${\cal
  C}^\infty({\cal N})$ with values in ${\cal C}^\infty(|\Omega|)$, where $(i_1,\cdots i_{2k})\in
[\![1,q]\!]^{2k}$ and $0\leq k\leq [q/2]$ ($[q/2]$ is the integer
part of $q/2$), by the relation
\[
\phi^*f = \sum_{k=0}^{[q/2]}\sum_{1\leq i_1<\cdots <i_{k}\leq
  q}a_{i_1\cdots i_{2k}}(f) \eta^{i_1}\cdots \eta^{i_{2k}}
= a_\emptyset(f) + \sum_{1\leq i_1<i_2\leq
  q}a_{i_1i_2}(f) \eta^{i_1}\eta^{i_2} + \cdots
\]
Here we will assume that the $a_{i_1\cdots i_{2k}}$'s are skew
symmetric in $(i_1,\cdots i_{2k})$. At this point it is useful to
introduce the following notations: For any positive integer $k$ we let $\I^q(k):=
\{(i_1,\cdots i_k)\in [\![1,q]\!]^k| i_1<\cdots <i_k\}$,
we denote by $I=(i_1,\cdots i_{k})$ an element of
$\I^q(k)$ and we then write $\eta^I:= \eta^{i_1}\cdots
\eta^{i_k}$. It will be also useful to use the convention $\I^q(0) =
\{\emptyset \}$.
We let $\I^q:= \cup_{k=0}^q\I^q(k)$, $\I^q_0:=
\cup_{k=0}^{[q/2]}\I^q(2k)$, $\I^q_1:=
\cup_{k=0}^{[(q-1)/2]}\I^q(2k+1)$
and $\I^q_2:= \cup_{k=1}^{[q/2]}\I^q(2k)$.
Hence the preceding relation can be
written
\begin{equation}\label{t*f}
\phi^*f = \sum_{k=0}^{[q/2]}\sum_{I\in\I^q(2k)} a_I(f)\eta^I =
\sum_{I\in \I^q_0} a_I(f)\eta^I
\end{equation}
or
\[
\forall x\in |\Omega|,\quad (\phi^*f)(x) = \sum_{I\in \I^q_0}
a_I(f)(x)\eta^I.
\]
{\bf Construction of morphisms}\\

\noindent
We start by providing a construction of morphisms satisfying
(\ref{ax+}) and (\ref{axprod}). We note $\pi: |\Omega|\times {\cal
  N}\longrightarrow {\cal N}$ the canonical projection map and
consider the vector bundle $\pi^*T{\cal N}$: 
the fiber over each point $(x,q)\in |\Omega|\times {\cal N}$ is the
tangent space $T_q{\cal N}$. For any $I\in \I^q_2$, we choose a
smooth section $\xi_I$ of $\pi^*T{\cal N}$ over $|\Omega|\times {\cal
  N}$ and we consider the $\R[\eta^1,\cdots ,\eta^q]_0$-valued vector field
\[
\Xi:= \sum_{I\in\I^q_2} \xi_I \eta^I.
\]
Alternatively $\Xi$ can be seen as a smooth family $\left(\Xi_x\right)_{x\in
|\Omega|}$ of smooth tangent vector fields  on ${\cal N}$ with
coefficients in $\R[\eta^1,\cdots ,\eta^q]_0$. So each $\Xi_x$ defines
a first order differential operator which acts on the algebra ${\cal 
  C}^\infty({\cal N})\otimes \Bbb{R}[\eta^1,\cdots ,\eta^q]_0$, i.e. the set of smooth
  functions on ${\cal N}$ with values in $\Bbb{R}[\eta^1,\cdots ,\eta^q]_0$, by the relation
\[
\forall f\in {\cal C}^\infty({\cal N})\otimes \Bbb{R}[\eta^1,\cdots ,\eta^q]_0, \quad
\Xi_x f = \sum_{I\in\I^q_2} ((\xi_I)_x\cdot f) \eta^I.
\]
Here we do not need to worry about the position of $\eta^I$ since it
is an even monomial. We now define (letting $\Xi^0 = 1$)
\[
e^\Xi := \sum_{n=0}^\infty {\Xi^n\over n!} = \sum_{n=0}^{[q/2]} {\Xi^n\over
  n!},
\]
which can be considered again as a smooth family parametrized by $x\in
|\Omega|$ of differential operators of order at
most $[q/2]$ acting on ${\cal C}^\infty({\cal N})\otimes
\Bbb{R}[\eta^1,\cdots ,\eta^q]_0$. Now we choose a smooth map $\varphi:|\Omega|
\longrightarrow {\cal N}$ and we consider the map
\[
\begin{array}{cccc}
1\times \varphi: & |\Omega| & \longrightarrow & |\Omega|\times {\cal N}\\
& x & \longmapsto & (x,\varphi(x))
\end{array}
\]
which parametrizes the graph of
$\varphi$. Lastly we construct
the following linear operator on ${\cal C}^\infty({\cal N})\subset {\cal
  C}^\infty({\cal N})\otimes \Bbb{R}[\eta^1,\cdots ,\eta^q]_0$:
\[
{\cal C}^\infty({\cal N})\ni f\longmapsto (1\times \varphi)^*\left(
  e^\Xi f\right) \in {\cal C}^\infty(\Omega),
\]
where
\[
\forall x\in |\Omega|,\quad
(1\times \varphi)^*\left( e^\Xi f\right)(x):= \left( e^{\Xi_x}
  f\right)(\varphi(x)) = \sum_{n=0}^{[q/2]} \left({(\Xi)_x^n\over
  n!}f\right)(\varphi(x)).
\]
We observe that actually, for any $x\in |\Omega|$, we only need to define $\Xi_x$ on
a neighbourhood of $\varphi(x)$ in ${\cal N}$, i.e. it suffices to
define the section $\Xi$ on a neighbourhood of the graph of $\varphi$
in $|\Omega|\times {\cal N}$ (or even on their Taylor expansion in $q$
at order $[q/2]$ around $\varphi(x)$).
\begin{lemm}\label{lemma1}
The map $f\longmapsto (1\times \varphi)^*\left( e^\Xi f\right)$ is a morphism from
${\cal C}^\infty({\cal N})$ to ${\cal C}^\infty(\Omega)_0$, i.e. satisfies
assumptions (\ref{ax+}) and (\ref{axprod}).
\end{lemm}
{\em Proof} --- Property (\ref{ax+}) is obvious, so we just need to prove (\ref{axprod}).
We first remark that, for any $x\in |\Omega|$, $\Xi_x$ satisfies the Leibniz rule:
\[
\forall f,g\in {\cal C}^\infty({\cal N})\otimes \Bbb{R}[\eta^1,\cdots,\eta^q]_0, \quad
\Xi_x (fg) = (\Xi_x f)g + f(\Xi_x g),
\]
which immediately implies by recursion that
\begin{equation}\label{leibnizc}
\forall f,g\in {\cal C}^\infty({\cal N})\otimes \Bbb{R}[\eta^1,\cdots,\eta^q]_0,\forall
n\in \N, \quad
\Xi_x^n(fg) = \sum_{j=1}^n {n!\over (n-j)!j!}(\Xi_x^{n-j}f)(\Xi_x^jg).
\end{equation}
We deduce easily that
\begin{equation}\label{expo}
\forall x\in |\Omega|,
\forall f,g\in {\cal C}^\infty({\cal N})\otimes \Bbb{R}[\eta^1,\cdots,\eta^q]_0, \quad
e^{\Xi_x} (fg) = \left(e^{\Xi_x} f\right) \left(e^{\Xi_x} g\right),
\end{equation}
by developping both sides and using (\ref{leibnizc}). Now relation
(\ref{expo}) is true in particular for functions $f,g\in {\cal
  C}^\infty({\cal N})$ and if we evaluate this identity at the point
$\varphi(x)\in {\cal N}$ we immediately conclude that $f\longmapsto (1\times
\varphi)^*\left( e^\Xi f\right)$ satisfies (\ref{axprod}).\bbox

\noindent The following result says that actually all morphisms are of
the previous type.
\begin{theo}\label{letheo} Let $\phi^*:{\cal C}^\infty({\cal N})\longrightarrow
  {\cal C}^\infty(\Omega)_0$ be a morphism. Then there exists a smooth
    map $\varphi:|\Omega|\longrightarrow {\cal N}$ and a smooth
    family $\left(\xi_I\right)_{I\in\I^q_2}$ of sections of $\pi^*T{\cal N}$ 
    defined on a neighbourhood of the graph of $\varphi$ in 
    $|\Omega|\times {\cal N}$, such that if $\Xi := \sum_{I\in\I^q_2}
    \xi_I \eta^I$, then
\begin{equation}\label{universal}
\forall f\in {\cal C}^\infty({\cal N}),\quad
\phi^*f = (1\times \varphi)^*\left( e^\Xi f\right).
\end{equation}
\end{theo}
{\em Proof} --- Let  $\phi^*:{\cal C}^\infty({\cal
    N})\longrightarrow {\cal C}^\infty(\Omega)_0$ which satisfies
(\ref{ax+}) and (\ref{axprod}). We denote by $a_I$ the functionals
involved in the identity (\ref{t*f}). We also introduce the following notation:
    for any $N\in \N$, ${\cal O}(\eta^{(N)})$ will represent a
    quantity of the form
\[
{\cal O}(\eta^{(N)}) = \sum_{n=N}^\infty \sum_{I\in
  \I^q(n)}c_I\eta^I,
\]
where the coefficients $c_I$'s may be real constants or functions. The
result will follow by proving by recursion on $n\in \N^*$ the following
property:
\begin{itemize}
\item
{\em
$(P_n)$: There exists a smooth map $\varphi:|\Omega|\longrightarrow
{\cal N}$ and there exists a
family of vector fields $\left(\xi_I\right)_I$, where $I\in \I^q(2k)$
and $1\leq k\leq n$, defined on a neighbourhood of the graph of
$\varphi$ in $|\Omega|\times {\cal N}$, such that if
\[
\Xi_n:= \sum_{k=1}^{n}\sum_{I\in\I^q(2k)}\xi_I\eta^I,
\]
then
\[
\forall f\in {\cal C}^\infty({\cal N}), \quad
\phi^*f = (1\times \varphi)^*\left( e^{\Xi_n} f\right) + {\cal
  O}\left(\eta^{(2n+1)}\right). 
\] }
\end{itemize}
{\em Proof of $(P_1)$} --- For start from relation
(\ref{axprod}) and
we expand both sides by using (\ref{t*f}): we first obtain by
identifying the terms of degree 0 in the $\eta^i$'s:
\[
\forall x\in |\Omega|, 
\forall f,g\in {\cal C}^\infty({\cal N}), \quad
a_\emptyset(fg)(x) = \left(a_\emptyset(f)(x)\right) \left(a_\emptyset(g)(x)\right),
\]
which implies that, for any $x\in |\Omega|$, there exists some value
$\varphi(x)\in {\cal N}$ such that
\[
\forall x\in |\Omega|, 
\forall f\in {\cal C}^\infty({\cal N}), \quad
a_\emptyset(f)(x) = f(\varphi(x)).
\]
In other words there exists a function $\varphi:
|\Omega|\longrightarrow {\cal N}$ such that $a_\emptyset(f) = f\circ
\varphi$. Since $a_\emptyset(f)$ must be ${\cal C}^\infty$ for any
smooth $f$, this implies that $\varphi\in {\cal
  C}^\infty(|\Omega|,{\cal N})$.
The relations between the terms of degree 2 in (\ref{axprod}) are:
$\forall x\in |\Omega|$, $\forall f,g\in {\cal C}^\infty({\cal N})$,
\[
\begin{array}{ccl}
\forall I\in \I^q(2), \quad
a_I(fg)(x)&  = & \left(a_I(f)(x)\right) \left(a_\emptyset(g)(x)\right) +
\left(a_\emptyset(f)(x)\right) \left(a_I(g)(x)\right)\\
&  = &
\left(a_I(f)(x)\right)g(\varphi(x)) + f(\varphi(x))\left(a_I(g)(x)\right),
\end{array}
\]
which implies that for any $x\in |\Omega|$, each $a_I(\cdot)(x)$ is a
derivation acting on ${\cal C}^\infty({\cal N})$, with support
$\{\varphi(x)\}$, i.e. $\forall I\in \I^q(2)$ there exist
tangent vectors  $\left(\xi_I\right)_x\in T_{\varphi(x)}{\cal N}$ such that 
\[
\forall f\in {\cal C}^\infty({\cal N}), \quad
a_I(f)(x) =  \left(\left(\xi_I\right)_x\cdot f\right)(\varphi(x))  .
\]
And since $a_I(f)$ must be smooth for any $f\in {\cal C}^\infty({\cal
  N})$, the vectors $\left(\xi_I\right)_x$ should depend smoothly on
  $x$, i.e. $x\longmapsto \left(\xi_I\right)_x$ is a smooth section of
  $\varphi^*T{\cal N}$. It is then possible (see the
  Proposition \ref{prop1} below) to extend it to a smooth section of
  $\pi^*T{\cal N}$ on a neighbourhood of the graph of $\varphi$. If we
  now set $\left(\Xi_1\right)_x:= \sum_{I\in
  \I^q(2)}\left(\xi_I\right)_x\eta^I$ we have on the one hand,
  $\forall x\in |\Omega|$, 
\[
\forall f\in {\cal N},\quad
e^{\left(\Xi_1\right)_x}f = f + \sum_{I\in
  \I^q(2)}\left(\left(\xi_I\right)_x\cdot f\right)\eta^I + {\cal 
  O}(\eta^{(3)})
\]
and on the other hand $\forall x\in |\Omega|$,
\[
(\phi^*f)(x) = f(\varphi(x)) + \sum_{I\in \I^q(2)}
  \left(\left(\xi_I\right)_x\cdot f\right)(\varphi(x))\eta^I + {\cal
  O}(\eta^{(3)}),
\]
from which $(P_1)$ follows.\\
{\em Proof of $(P_n)\Longrightarrow (P_{n+1})$} --- We assume $(P_n)$
so that a map $\varphi\in {\cal C}^\infty(|\Omega|,{\cal N})$ and a
vector field $\Xi_n$ have been constructed. Let us denote by
$b_I$ the linear forms on ${\cal C}^\infty({\cal N})$ such that 
\begin{equation}\label{expdn}
(1\times \varphi)^*\left(e^{\Xi_n}f\right) = \sum_{k=0}^{[q/2]} \sum_{I\in \I^q(2k)}
b_I(f)\eta^I.
\end{equation}
Then property $(P_n)$ is equivalent to
\begin{equation}\label{pn}
\forall k\in [\![0,n]\!],\forall I\in \I^q(2k),\quad a_I = b_I.
\end{equation}
We use Lemma \ref{lemma1}: it says us that $f\longmapsto (1\times
\varphi)^*\left(e^{\Xi_n}f\right)$ is a morphism, hence
$(1\times \varphi)^*\left(e^{\Xi_n}(fg)\right) = \left[(1\times
\varphi)^*\left(e^{\Xi_n}f\right)\right] \left[(1\times
\varphi)^*\left(e^{\Xi_n}g\right)\right]$,
so by using (\ref{expdn}):
\begin{equation}\label{bn+1}
\sum_{k=0}^{n+1} \sum_{I\in \I^q(2k)}b_I(fg)\eta^I =
\sum_{k=0}^{n+1} \sum_{j=0}^k\sum_{J\in \I^q(2k-2j),K\in \I^q(2j)}
b_J(f)b_K(g) \eta^J\eta^K + {\cal O}(\eta^{(2n+3)}).
\end{equation}
But the morphism property (\ref{axprod}) for $\phi^*$ implies also
\begin{equation}\label{an+1}
\sum_{k=0}^{n+1} \sum_{I\in \I^q(2k)}a_I(fg)\eta^I =
\sum_{k=0}^{n+1} \sum_{j=0}^k\sum_{J\in \I^q(2k-2j),K\in \I^q(2j)}
a_J(f)a_K(g) \eta^J\eta^K + {\cal O}(\eta^{(2n+3)}).
\end{equation}
We now substract (\ref{bn+1}) to (\ref{an+1}) and use (\ref{pn}): it
gives us
\[
\sum_{I\in \I^q(2n+2)}\left( a_I(fg)-b_I(fg)\right) \eta^I = 
\sum_{I\in \I^q(2n+2)}\left[\left( a_I(f)-b_I(f)\right)a_\emptyset(g) +
a_\emptyset(f) \left( a_I(g)-b_I(g)\right)\right] \eta^I.
\]
Hence if we denote $\delta a_I:= a_I - b_I$, we obtain that
\[
\forall I\in \I^q(2n+2),\quad
\delta a_I(fg) = \delta a_I(f)(g\circ \varphi) + (f\circ \varphi)\delta a_I(g).
\]
By the same reasoning as in the proof of $(P_1)$, we conclude that,
$\forall I\in \I^q(2n+2)$, there exist
smooth sections $\xi_I$ of $\pi^*T{\cal N}$ defined on a neighbourhood of the graph of $\varphi$, 
such that
\[
\forall x\in |\Omega|,\forall I\in \I^q(2n+2),\quad
\delta a_I(f)(x) = \left(\left(\xi_I\right)_x\cdot f\right)(\varphi(x)).
\]
Now let us define
\[
\Xi_{n+1}:= \Xi_n + \sum_{I\in \I^q(2n+2)}\xi_I\eta^I.
\]
Then it turns out that
\[
\begin{array}{ccl}
\displaystyle e^{\Xi_{n+1}}f & = &
      \displaystyle \sum_{k=0}^{n+1} 
      {\left(\Xi_n+\sum_{I\in \I^q(2k+2)}\xi_I\eta^I \right)^k\over
      k!}f + {\cal O}(\eta^{(2n+3)})\\
 & = & \displaystyle  \sum_{k=0}^{n+1} {\Xi_n^k\over k!}f
      + \sum_{I\in \I^q(2n+2)}\xi_I\cdot f\eta^I + {\cal
      O}(\eta^{(2n+3)})\\ 
 & = &\displaystyle  e^{\Xi_n}f + \sum_{I\in
      \I^M(2n+2)}\xi_I\cdot f\eta^I + {\cal O}(\eta^{(2n+3)}),
\end{array}
\]
so that
\[
(1\times \varphi)^*\left(e^{\Xi_{n+1}}f\right) = \phi^*f + {\cal
  O}(\eta^{(2n+3)}).
\]
Hence we deduce $(P_{n+1})$.\bbox

\begin{prop}\label{prop1}
In the preceding result, it is possible to construct smoothly the vector fields
$\xi_I$'s in such a way that, $\forall x\in |\Omega|$,
\[
\forall I,J\in \I^q_2,\quad
\left[\left(\xi_I\right)_x,\left(\xi_J\right)_x\right] = 0.
\]
\end{prop}
{\em Proof} --- Recall that in the previous proof, in order to build
$\Xi_{n+1}$ out of $\Xi_n$, we introduced, for each $I\in \I^q(2n+2)$,
an unique smooth section $x\longmapsto \left(\xi_I\right)_x$ of
$\varphi^*T{\cal N}$. We will explain here how
to extend each such vector fields defined along the graph of $\varphi$
to a neighbourhood of the graph of $\varphi$ in $|\Omega|\times {\cal
  N}$ in order to achieve the claim in the proposition.
For that purpose we prove that for some set
\[
{\cal V}:= \{
(x,\xi,q)\in \varphi^*T{\cal N} \times {\cal N}|\ x\in
|\Omega|, \xi\in T_{\varphi(x)}{\cal N}, q\in V_{\varphi(x)}\},
\] 
where each $V_{\varphi(x)}$ is a neighbourhood of $\varphi(x)$ in
${\cal N}$, there exists a smooth map
\[
\begin{array}{ccc}
{\cal V} & \longrightarrow & T{\cal N}\\
(x,\xi,q) & \longmapsto & (q, \Bbb{V}(x,\xi,q))
\end{array}
\]
such that $\forall (x,\xi)\in \varphi^*T{\cal N}$,
$\Bbb{V}(x,\xi,\varphi(x)) = \xi$ and $\forall x\in |\Omega|$
fixed, $\forall \xi,\zeta\in
T_{\varphi(x)}{\cal N}$,
$[\Bbb{V}(x,\xi,\cdot),\Bbb{V}(x,\zeta,\cdot)] = 0$, i.e. the vector
fields $q\longmapsto \Bbb{V}(x,\xi,q)$ and $q\longmapsto
\Bbb{V}(x,\zeta,q)$ commute on $V_{\varphi(x)}$. Then the proposition
will follows by extending each vector $\left(\xi_I\right)_x\in T_{\varphi(x)}{\cal
  N}$ on $V_{\varphi(x)}$ by $q\longmapsto \Bbb{V}(x,\left(\xi_I\right)_x,q)$.\\

\noindent
The construction is the following. Let $\left(U_a\right)_{a\in A}$
be a covering of ${\cal N}$ by open subsets, let
$\left(\chi_a\right)_{a\in A}$ be a partition of unity and
$\left(y_a\right)_{a\in A}$ be a family of charts associated with this
covering. For any $x\in |\Omega|$, let $A_x:= \{a\in A|\ \varphi(x)\in
U_a\}$. For any $a\in A_x$ and for any linear isomorphism $\ell: T_{\varphi(x)}{\cal
  N}\longrightarrow \R^n$, where $n = \hbox{dim}{\cal N}$, let
$R_{x,\ell,a}$ be the unique linear automorphism of $\R^n$ such that
\[
R_{x,\ell,a}\circ dy_{a|\varphi(x)} = \ell.
\]
We then set
\[
\forall q\in {\cal N},\quad
y_{x,\ell}(q):= \sum_{a\in A_x} \chi_a(q) R_{x,\ell,a}\circ y_a(q).
\]
We observe that $dy_{x,\ell|\varphi(x)} = \ell$ and hence, by the
inverse mapping theorem, there exists an open neighbourhood $V_{\varphi(x)}$ of
$\varphi(x)$ in ${\cal N}$ such that the restriction of $y_{x,\ell}$
to $V_{\varphi(x)}$ is a diffeomorphism. We then define
\[
\forall q\in V_{\varphi(x)},\quad
\Bbb{V}(x,\xi,q):= \left(dy_{x,\ell |q}\right)^{-1} \left(
  \ell(\xi)\right).
\]
Because of the obvious relation $y_{x,u\circ \ell} = u\circ
y_{x,\ell}$ for all linear automorphism $u$ of $\R^n$, it is clear that the
definition of $\Bbb{V}(x,\xi,q)$ does not depend on $\ell$ (for the
same reason $V_{\varphi(x)}$ is also independant of $\ell$). Moreover
$q\longmapsto \Bbb{V}(x,\xi,q)$ is simply a vector field which is a
linear combination with constant coefficients of the vector fields
$\left({\partial \over \partial y_{x,\ell}^i}\right)_{i=1,\cdots ,n}$ so that the
property $[\Bbb{V}(x,\xi,\cdot),\Bbb{V}(x,\zeta,\cdot)] = 0$ follows.
Note also that these vector fields are of course not canonical
since they obviously depend on the charts. \bbox

\begin{rema}\label{remaxixiy=0}
If we assume furthermore that the image of $\varphi$ is contained in an
open subset $U$ of ${\cal N}$ such that there exists a local chart $y
= (y^1,\cdots ,y^n):U\longrightarrow \R^n$, then it is possible to
choose all the vector fields $\xi_I$ such that
\begin{equation}\label{xixiy=0}
\forall x\in |\Omega|, \forall I,J\in \I^q_2,\quad 
\left(\xi_I\right)_x \cdot \left(\xi_J\right)_x \cdot y = 0.
\end{equation}
Indeed in this case the proof of Proposition \ref{prop1} is much
simpler, since we do not need to use a partition of unity in order to
build $\Bbb{V}$. We just set ${\cal V}:= \{
(x,\xi,q)\in \varphi^*T{\cal N} \times {\cal N}|\ x\in
|\Omega|, \xi\in T_{\varphi(x)}{\cal N}, q\in U\}$ and define
$\Bbb{V}$ by $\Bbb{V}(x,\xi,q):= (dy_{|q})^{-1}\circ
dy_{|\varphi(x)}(\xi)$. Then for each $(x,\xi)\in \varphi^*T{\cal N}$
fixed, the vector field $q\longmapsto
\Bbb{V}(x,\xi,q)$ has constant coordinates in the variables
$y^\alpha$. Hence (\ref{xixiy=0}) follows.
\end{rema}
\begin{rema}
We can write an alternative formula for $e^\Xi $ by developping this
exponential: in each term of the form $\left(\sum_I
  \xi_I\eta^I\right)^n$ we can see that each monomial which
appears contains at most one time any operator $\xi_I$, so we obtain
\begin{equation}\label{devexp}
e^\Xi  = \sum_{I\in\I^q_0}\eta^I\left(
\sum_{n\geq 0}{1\over n!}
\sum_{I_1,\cdots ,I_n\in \I^q_0}
\epsilon^{I_1\cdots I_n}_I \xi_{I_1}\cdots \xi_{I_n}\right),
\end{equation}
with the convention that the $\I^q_0(0) = \emptyset$ contribution
is the identity.
Here we have introduced the notation $\epsilon^{I_1\cdots I_n}_I$:
first all the $\epsilon^{I_1\cdots I_n}_\emptyset$'s vanish except for
$\epsilon^\emptyset_\emptyset = 1$, so that $e^\Xi = 1\ \hbox{\em 
mod}[\eta^1,\cdots ,\eta^q]$. Second, for $k\geq 1$, if $I_1 = (i_{1,1},\cdots
,i_{1,2k_1})$, $\cdots$, $I_n = (i_{n,1},\cdots ,i_{n,2k_n})$ and
$I=(i_1,\cdots ,i_{2k})$, we write that $I_1\sqcup \cdots \sqcup I_n
=I$ if and only if $k_1+\cdots +k_n = k$, $\{i_{1,1},\cdots ,i_{1,2k_1},\cdots
,i_{n,1},\cdots ,i_{n,2k_n}\} = \{i_1,\cdots ,i_{2k}\}$ and $\forall
j$, $I_j\neq \emptyset$ (i.e. $k_j>0$). Then
\begin{itemize}
\item if $I_1\sqcup \cdots \sqcup I_n
  \neq I$, $\epsilon^{I_1\cdots I_n}_I = 0$ 
\item if $I_1\sqcup \cdots \sqcup I_n
  = I$, $\epsilon^{I_1\cdots I_n}_I$ is the signature of the
permutation\\
$(i_{1,1},\cdots ,i_{1,2k_1},\cdots ,i_{n,1},\cdots
,i_{n,k_{2n}}) \longmapsto (i_1,\cdots ,i_{2k})$.
\end{itemize}
The preceding expression of $e^\Xi $ can be recovered by another way: since all the operators
$\eta^I\xi_I$ commute, we have
\[
e^\Xi  = e^{\sum_{I\in \I^q_2} \eta^I\xi_I} = \prod_{I\in \I^q_2} e^{\eta^I\xi_I}
= \prod_{I\in \I^q_2} (1 + \eta^I\xi_I),
\]
which gives also the same result by a straightforward development.
\end{rema}

\section{A factorization of the morphism $\phi^*$}

\subsection{Integrating the vector fields $\xi_I$'s}
In the same spirit as a tangent vector at a point $q$ to a manifold
${\cal N}$ can be seen as the time derivative of a smooth curve which
reaches $q$ we can describe the $\eta^I$-components of the morphism $\phi^*$ as
higher order approximations of a smooth map from some vector space
with values in
${\cal N}$. Indeed let $\phi^* \in \hbox{Mor}({\cal C}^\infty({\cal
  N}),{\cal C}^\infty(\Omega)_0)$: then
by the preceding result $\phi^*$ is characterized by a map
$\varphi\in {\cal C}^\infty(|\Omega|,{\cal N})$ and $2^{q-1}-1$ vector
fields\footnote{note that $\hbox{card}
  \I^q(2k) = {q!\over (q-2k)!(2k)!}$ and $\sum_{k=0}^{[q/2]}{q!\over
(q-2k)!(2k)!} = 2^{q-1}$} $\xi_I$ tangent to ${\cal N}$ defined on a neighbourhood of the
graph of $\varphi$ in $|\Omega|\times {\cal N}$. By proposition
\ref{prop1} these vector
fields can moreover be chosen so that they pairwise commute when $x\in
|\Omega|$ is fixed. So,
for any $x\in |\Omega|$ we can integrate simultaneously all vector
fields $\left(\xi_I\right)_x$ in order to construct a map 
\[
\Phi(x,\cdot): U_x(\Lambda^{2*}_+\Bbb{R}^q) \longrightarrow {\cal N},
\]
where $\Lambda^{2*}_+\Bbb{R}^q \simeq \R^{2^{q-1}-1}$ is the subspace
of even elements of positive degree of the exterior algebra
$\Lambda^*\Bbb{R}^q$ and $U_x(\Lambda^{2*}_+\Bbb{R}^q)$ is a
neighbourhood of 0 in $\Lambda^{2*}_+\Bbb{R}^q$, such that
\begin{equation}\label{condinitial}
\Phi(x,0) = \varphi(x) 
\end{equation}
and, denoting by $\left(\x^I\right)_{I\in \I^q_2}$ the linear
coordinates on $\Lambda^{2*}_+\Bbb{R}^q$,
\begin{equation}\label{dux}
{\partial \Phi\over   \partial \x^I}(x,\x)
= \xi_I(\Phi(x,\x))\quad\forall \x\in  U_x(\Lambda^{2*}_+\Bbb{R}^q), \forall I\in \I^q_2.
\end{equation}
We hence obtain a map $\Phi$ from a neighbourhood of $|\Omega|\times
\{0\}$ in $|\Omega|\times \Lambda^{2*}_+\Bbb{R}^q$ to ${\cal N}$.
By using a cut-off function argument we can extend this map to
an application $\Phi: |\Omega|\times
\Lambda^{2*}_+\Bbb{R}^q\longmapsto {\cal N}$. Lastly we introduce the
$\R[\eta^1,\cdots ,\eta^q]$-valued vector field on $|\Omega|\times \Lambda^{2*}_+\Bbb{R}^q$
\[
\vartheta:= \sum_{I\in \I^q_2}\eta^I{\partial \over  \partial \x^i},
\]
so that by (\ref{dux}) $\Phi_*\vartheta = \Xi = \sum_{I\in \I^q_2}\eta^I\xi_I$. Then relation
(\ref{universal}) implies
\[
\forall f\in {\cal C}^\infty({\cal N}),
\forall x\in |\Omega|,\quad  \phi^*f(x) = \left( e^\vartheta (f\circ
  \Phi)\right) (x,0)
\]
or by letting $\iota: |\Omega| \longrightarrow |\Omega|\times
\Lambda^{2*}_+\Bbb{R}^q$, $x\longmapsto (x,0)$ to be the canonical
injection,
\begin{equation}\label{phistarf}
\phi^*f = \iota^*\left( e^\vartheta (f\circ
  \Phi)\right).
\end{equation}
Alternatively by using (\ref{devexp}) we have
\begin{equation}\label{udonnet}
\forall x\in |\Omega|,\quad
\phi^*f(x) = \sum_{I\in\I^q_0}\eta^I \left(
\sum_{k\geq 0}{1\over k!} \sum_{I_1, \cdots
     ,I_k \in \I^q_0} \epsilon^{I_1\cdots I_k}_I
{\partial^k(f\circ \Phi)\over \partial \x^{I_1}\cdots \partial \x^{I_k}}(x,0)
\right).
\end{equation}
It is useful to
introduce the differential operators ${\cal D}_\emptyset:= 1$ and
\[
{\cal D}_I:= \sum_{k\geq 0} {1\over k!}
\sum_{I_1, \cdots ,I_k \in \I^q_0}\epsilon^{I_1\cdots I_k}_I
{\partial^k\over \partial \x^{I_1}\cdots \partial \x^{I_k}},
\]
so that $\phi^*f(x) = \sum_{I\in\I^q_0}\eta^I{\cal D}_I(f\circ \Phi)(x,0)$.
Conversely to any map smooth map $\Phi\in {\cal
  C}^\infty(|\Omega|\times \Lambda^{2*}_+\R^q,{\cal N})$ we can associate a unique
  morphism $\phi^*\in \hbox{Mor}({\cal C}^\infty({\cal
    N}),\Bbb{R}[\eta^1,\cdots ,\eta^q]_0)$
  defined by (\ref{phistarf}) or (\ref{udonnet}). This defines an application
\[
\begin{array}{ccc}
{\cal C}^\infty(|\Omega|\times \Lambda^{2*}_+\R^q,{\cal N}) & \longrightarrow &
\hbox{Mor}({\cal C}^\infty({\cal N}),{\cal C}^\infty(\Omega)_0) \\
\Phi & \longmapsto & \Phi^*_{|\circ},
\end{array}
\]
where $\forall f\in {\cal C}^\infty({\cal N})$, $\Phi^*_{|\circ}f = \iota^*\left( e^\vartheta (f\circ
  \Phi)\right)$.
It is clear from the previous discussion that this application is
onto. It is however certainly not injective, since $\Phi^*_{|\circ}$ depends
only on the $[q/2]$-th order Taylor expansion of $\Phi$ at $0$. This will
be precised in the following.

\subsection{Expressions using local coordinates on the target
  manifold}
Assume that we have local coordinates on ${\cal N}$:
we let $U$ to be
an open subset of ${\cal N}$ and we consider a chart $y =
(y^1,\cdots ,y^n): U \longrightarrow V\subset \R^n$. 
Then any function $f:U \longrightarrow \R$ can be represented by an
unique function $F:V\longrightarrow \R$ such that $f = F\circ
y$. For any $y_0\in V\subset \R^n$ let $P^{[q/2]}_{F,y_0}$ be the $[q/2]$-th order Taylor
expansion of $F$ at $y_0$ and $R^{[q/2]}_{F,y_0}$ be the rest, so that we have
the decomposition $F(y) = P^{[q/2]}_{F,y_0}(y) + 
R^{[q/2]}_{F,y_0}(y)$. The expressions for $P^{[q/2]}_{F,y_0}$ and
$R^{[q/2]}_{F,y_0}$ are:
\[
\forall y\in \R^n,\quad
P^{[q/2]}_{F,y_0}(y) = \sum_{r\in \N^n,|r|\leq [q/2]} {\partial
  ^rF\over (\partial y)^r}(y_0){(y-y_0)^r\over r!}
\]
and
\[
\forall y\in V,\quad
R^{[q/2]}_{F,y_0}(y) = \sum_{r\in \N^n,|r|= [q/2]+1}
(y-y_0)^rR_{F,y_0,r}(y),
\]
where, if $r =(r_1,\cdots ,r_n)\in \N^n$, $|r|:= r_1+ \cdots +r_n$,
$(y)^r:= (y^1)^{r_1}\cdots (y^n)^{r_n}$ and ${\partial ^rF\over (\partial
  y)^r}:= {\partial ^{|r|}F\over (\partial y^1)^{r_1}\cdots (\partial
y^n)^{r_n}}$, assuming that $V$ is star-shaped around $y_0$,
\[
R_{F,y_0,r}(y):= {[q/2]+1\over r!}\int_0^1 (1-t)^{[q/2]} {\partial ^rF\over
  (\partial y)^r}(y_0+t(y-y_0))dt.
\]
\begin{prop}\label{proptaylor}
Let $y:{\cal N}\supset U\longrightarrow V\in \R^n$ be a local chart
  and $\phi^*:{\cal C}^\infty(U)\longrightarrow {\cal
  C}^\infty(\Omega)_0$ be a morphism. For any $f\in {\cal
  C}^\infty(U)$ let $F\in {\cal C}^\infty(V)$ such that $f = F\circ
  y$. Then $\forall x_0\in |\Omega|$,
\begin{equation}\label{taylor}
(\phi^*f)(x_0) = \sum_{r\in \N^n,|r|\leq [q/2]} {\partial
  ^rF\over (\partial y)^r}(y_0){(\phi^*y-y_0)^r\over r!},
\end{equation}
where $y_0$ is the unique point in $\R^n$ such that $y\circ
\phi(x_0)-y_0$ has nilpotent components.
\end{prop}
{\em Proof} --- The morphism property implies that 
\begin{equation}\label{taylorrest}
\phi^*(F\circ y) = \phi^*\left(P^{[q/2]}_{F,y_0}(y)\right) + 
\sum_{r\in \N^n,|r|= [q/2]+1}
\phi^*\left((y-y_0)^r \right)
\phi^*\left(R_{F,y_0,r}\circ y\right).
\end{equation}
But still by using the morphism property we have
$\phi^*\left(P(y)\right) = P\left(\phi^*y\right)$ for any polynomial
$P$ in $n$ real variables. Hence 
\[
\phi^*f = \phi^*(F\circ y) = P^{[q/2]}_{F,y_0}\left(\phi^*y\right) + 
\sum_{r\in \N^n,|r|= [q/2]+1}
\left(\phi^*y-y_0 \right)^r
\phi^*\left(R_{F,y_0,r}\circ y\right).
\]
In particular when we evaluate this last identity at
the point $x_0$ we get (\ref{taylor}) because $\left(\phi^*y-y_0
\right)^r(x_0) = 0$ for $|r|= [q/2]+1$. \bbox

\noindent
Now let $\Phi: |\Omega|\times \Lambda^{2*}_+\R^q
\longrightarrow U\subset {\cal N}$, then 
we have the diagram: \xymatrix{|\Omega|\times
  \Lambda^{2*}_+\R^q \ar@{->}[r]^\Phi \ar@{->}[dr]^{y\circ \Phi}
& {\cal N}\supset U \ar@{->}[d]^y \ar@{->}[r]^f & \R \\
& \R^n\supset V \ar@{->}[ur]^F & }
\begin{coro}\label{prop2.1}
Let $y:U \longrightarrow \R^n$ be a local
  chart on ${\cal N}$ and let $\Phi,\widetilde{\Phi}\in {\cal C}^\infty(|\Omega|\times
  \Lambda^{2*}_+\R^q,{\cal N})$ such that $\iota^*\Phi =
  \iota^*\widetilde{\Phi} =:\varphi$. Then
\begin{equation}\label{mu=mutilde}
\Phi^*_{|\circ} = \widetilde{\Phi}^*_{|\circ}
\end{equation}
if and only if
\begin{equation}\label{u=utilde}
\forall \alpha,  \forall I\in \I^q_0,\forall x\in |\Omega|,\quad
{\cal D}_I(y^\alpha\circ\Phi)(x,0) = {\cal D}_I(y^\alpha\circ\widetilde{\Phi})(x,0).
\end{equation}
\end{coro}

\noindent
{\em Proof} --- Since $\Phi^*_{|\circ}f = \iota^*\sum_{I\in
  \I^q_0}\eta^I{\cal D}_I(f\circ \Phi)$ condition (\ref{u=utilde})
  just means that $\forall \alpha$, $\Phi^*_{|\circ}y^\alpha =
  \widetilde{\Phi}^*_{|\circ} y^\alpha$ and hence is a trivial
  consequence of (\ref{mu=mutilde}). Conversely if (\ref{u=utilde}) is
  true then we recover (\ref{mu=mutilde}) by applying (\ref{taylor})
  for $\phi^* = \Phi^*_{|\circ}$ and $\phi^* =
  \widetilde{\Phi}^*_{|\circ}$ and with $y_0 = \varphi(x_0)$. \bbox

\noindent
It is natural to define the following equivalence relation in
${\cal C}^\infty(|\Omega|\times \Lambda^{2*}_+\R^q,{\cal N})$: for any
$\Phi,\widetilde{\Phi}\in {\cal C}^\infty(|\Omega|\times \Lambda^{2*}_+\R^q,{\cal N})$ 
\[
\Phi\sim \widetilde{\Phi} \quad \Longleftrightarrow \quad
\Phi^*_{|\circ} = \widetilde{\Phi}^*_{|\circ}.
\]
Then clearly morphisms in $\hbox{Mor}({\cal C}^\infty({\cal
  N}),{\cal C}^\infty(\Omega)_0)$ are in one to one correspondence with 
equivalence classes in ${\cal C}^\infty(|\Omega|\times \Lambda^{2*}_+\R^q,{\cal
  N})/\sim$. This gives us a direct 
geometric picture (which we shall discuss below) of a map
  $\phi:\R^{p|q}\supset \Omega\longrightarrow {\cal N}$
(thought as dual to a morphism $\phi^*$ in $\hbox{Mor}({\cal
  C}^\infty({\cal N}),{\cal C}^\infty(\Omega)_0)$): it 
can be identified with a class of maps in ${\cal
  C}^\infty(|\Omega|\times \Lambda^{2*}_+\R^q,{\cal N})/\sim$, i.e. a map 
into ${\cal N}$ surrounded by a family of infinitesimal deformations
inside ${\cal N}$. 

\subsection{The chain rule for the operators ${\cal D}_I$}
We exploit relation (\ref{taylor}) again but we use a different
expression for the Taylor polynomial 
\[
P^{[q/2]}_{F,y_0}(y) =  \sum_{k=0}^{[q/2]} {1\over k!}
  \sum_{\alpha_1,\cdots , \alpha_k = 1}^n {\partial^kF\over \partial
  y^{\alpha_1}\cdots \partial  y^{\alpha_k}}(y_0)
  (y^{\alpha_1}-y^{\alpha_1}_0) \cdots
  (y^{\alpha_k} - y^{\alpha_k}_0).
\]
Hence by (\ref{taylor})
\begin{equation}\label{taylor2}
\Phi^*_{|\circ}f(x_0) =  \sum_{k=0}^{[q/2]} {1\over k!}
  \sum_{\alpha_1,\cdots , \alpha_k = 1}^n {\partial^kF\over \partial
  y^{\alpha_1}\cdots \partial  y^{\alpha_k}}(y_0) \prod_{\ell=1}^k
  \left(\Phi^*_{|\circ}y^{\alpha_\ell} -y_0^{\alpha_\ell}\right).
\end{equation}
But since
\[
\Phi^*_{|\circ}y^{\alpha}(x_0) -y^{\alpha}_0= \sum_{I\in\I^q_2}\eta^I
{\cal D}_I(y^{\alpha}\circ \Phi)(x_0,0),
\]
we deduce by a substitution
\[
\Phi^*_{|\circ}f(x_0) = F(y_0) + \sum_{k=1}^{[q/2]}
{1\over k!} \sum_{I;I_1, \cdots ,I_k\in \I^q_0} \eta^I  \epsilon^{I_1\cdots
  I_k}_I\sum_{\alpha_1,\cdots , \alpha_k = 1}^n {\partial^kF\over \partial
  y^{\alpha_1}\cdots \partial  y^{\alpha_k}}(y_0) \prod_{\ell=1}^k
{\cal D}_{I_\ell}(y^{\alpha_\ell} \circ \Phi)(x_0,0).
\]
But on the other hand we have
\[
\Phi^*_{|\circ}f(x_0) =  f\circ \varphi(x_0) + \sum_{I\in\I^q_2}\eta^I {\cal D}_I
(f\circ \Phi)(x_0,0) = F(y_0) + \sum_{I\in\I^q_2}\eta^I {\cal D}_I (F\circ
y\circ \Phi)(x_0,0).
\]
These two relations give us by an identification an expression for each ${\cal D}_I (F\circ
y\circ \Phi)(x_0,0)$ in terms of ${\cal D}_{I}(y^{\alpha} \circ
\Phi)(x_0,0)$. By setting $Y^\alpha:= y^\alpha\circ \Phi$ it can be
formulated as follows
\begin{prop}\label{propchainrule}
For any map $Y\in {\cal C}^\infty(|\Omega|\times
\Lambda^{2*}_+\R^q,\R^n)$, for any $x_0\in |\Omega|$, for any open
neighbourhood $V$ of $y_0:= Y(x_0,0)$ in $\R^n$ and for
any map $F\in{\cal C}^\infty(V)$, we have $\forall I\in \I^q_2$,
\begin{equation}\label{2.2.3}
{\cal D}_I(F\circ Y)(x_0,0) = \sum_{k\geq 0}{1\over k!}\sum_{I_1, \cdots ,
      I_k\in \I^q_0} 
      \epsilon^{I_1\cdots I_k}_I \sum_{\alpha_1,\cdots ,\alpha_k =
      1}^n {\partial ^kF\over \partial y^{\alpha_1}\cdots
      \partial y^{\alpha_k}}(y_0)\prod_{\ell=1}^k 
{\cal D}_{I_\ell}Y^{\alpha_\ell}(x_0,0).
\end{equation}
\end{prop}

\noindent
{\bf An application}\\

\noindent
We use a specialization of the identity (\ref{2.2.3}) by choosing
  $\R^n = \Lambda^{2*}_+\R^q$, and by substituting to $Y$ a smooth map
  $S:\Lambda^{2*}_+\R^q \longrightarrow  \Lambda^{2*}_+\R^q$ such that
  $S(0) = 0$. We hence get
\[
{\cal D}_I(F\circ S)(0) = \sum_{p\geq 0}{1\over p!}\sum_{I_1,
      \cdots ,I_p\in \I^q_0} 
      \epsilon^{I_1\cdots I_p}_I \sum_{J_1,\cdots ,J_p\in \I^q_0}
{\partial ^pF\over \partial \x^{J_1}\cdots
      \partial \x^{J_p}}(0)
\left({\cal D}_{I_1}S^{J_1}\cdots {\cal D}_{I_p}S^{J_p}\right)(0).
\]
In the special case where ${\cal D}_IS^J(0) = \delta_I^J$ this
simplifies to
\begin{equation}\label{2.2.4}
{\cal D}_I(F\circ S)(0) = \sum_{p\geq 0}{1\over p!}\sum_{I_1,
      \cdots ,I_p\in \I^q_0} 
      \epsilon^{I_1\cdots I_p}_I  {\partial ^pF\over \partial \x^{I_1}\cdots
      \partial \x^{I_p}}(0) = {\cal D}_IF(0).
\end{equation}
We conclude that if $S:\Lambda^{2*}_+\R^q\longrightarrow \Lambda^{2*}_+\R^q$
is a smooth diffeomorphism such that $S(0) = 0$ and ${\cal D}_IS^J(0)
= \delta_I^J$, then $V\sim V\circ S$. Hence if we define
\[
{\cal T}_q:=\{
\hbox{diffeomorphisms } S:\Lambda^{2*}_+\R^q\longrightarrow \Lambda^{2*}_+\R^q
| S(0) = 0, {\cal D}_IS^J(0)= \delta_I^J\}
\]
then we remark that ${\cal T}_q$ is a group for the composition law (another consequence of
(\ref{2.2.4})) and we see that the morphism $\Phi^*_{|\circ}$ is characterized
by the behaviour of $\Phi$ modulo the action of ${\cal T}_q$ hence by
duality we can identify a map $T:\R^{0|q}\longrightarrow {\cal N}$
with a class of maps from $\Lambda^{2*}_+\R^q$ to ${\cal N}$ modulo the
action of ${\cal T}_q$ on $\Lambda^{2*}_+\R^q$.

\subsection{Leibniz identities for the operators ${\cal D}_I$}
The operators ${\cal D}_I$ satisfy nice Leibniz type
identities:
\begin{prop}\label{propLeib}
For any pair of functions $a,b\in {\cal C}^\infty(|\Omega|\times \Lambda^{2*}_+\R^q)$ and 
for any $I\in \I^q_0$, 
\begin{equation}\label{leibniz}
{\cal D}_I(ab) = \sum_{I_1, I_2\in \I^q_0} \epsilon^{I_1 I_2}_I
\left({\cal D}_{I_1}a\right)
\left({\cal D}_{I_2}b\right), 
\end{equation}
where in the summation we allow $(I_1,I_2) = (\emptyset,I)$ or
$(I,\emptyset)$.
\end{prop}
{\em Proof} --- By applying relation (\ref{expo}) for $\vartheta:=
\sum_I\eta^I{\partial \over \partial \x^I}$ we obtain
\begin{equation}\label{ab}
\forall a,b\in {\cal C}^\infty(|\Omega|\times \Lambda^{2*}_+\R^q),\quad
e^\vartheta(ab) = \left(e^\vartheta a\right)
\left(e^\vartheta b\right).
\end{equation}
And by using $e^\vartheta a = \sum_{I\in \I^q_0} \eta^I\left({\cal D}_Ia\right)$
to developp this relation we obtain (\ref{leibniz}).\bbox

\noindent
A straightforward consequence of Proposition \ref{propLeib} is that the set
\[
{\cal I}^q(|\Omega|):= \{f\in {\cal C}^\infty(|\Omega|\times \Lambda^{2*}_+\R^q)|
 \forall I\in \I^q_0, \iota^*\left({\cal D}_If\right) = 0\} 
\]
is an ideal of the commutative algebra $\left({\cal
    C}^\infty(|\Omega|\times \Lambda^{2*}_+\R^q),+,\cdot\right)$. Hence the quotient
${\cal A}^q(|\Omega|):= {\cal C}^\infty(|\Omega|\times \Lambda^{2*}_+\R^q)/{\cal I}^q(|\Omega|)$ is an
algebra over $\R$. We will recover that this algebra is isomorphic to
${\cal C}^\infty(|\Omega|)[\eta^1,\cdots ,\eta^q]_0$. First we may
    also write ${\cal A}^q(|\Omega|) \simeq {\cal
    C}^\infty_{pol}(|\Omega|\times \Lambda^{2*}_+\R^q)/{\cal I}^q_{pol}(|\Omega|)$, where ${\cal
    C}^\infty_{pol}(|\Omega|\times \Lambda^{2*}_+\R^q)$ is the subalgebra of smooth
    functions on $|\Omega|\times
    \Lambda^{2*}_+\R^q$ which have a polynomial dependence in the
    variables $\x^I$ and
    ${\cal I}^q_{pol}(|\Omega|) = {\cal C}^\infty_{pol}(|\Omega|\times \Lambda^{2*}_+\R^q) \cap
    {\cal I}^q(|\Omega|)$. And any function $f\in {\cal
    C}^\infty_{pol}(|\Omega|\times \Lambda^{2*}_+\R^q)$ can be written
\[
f(x,\x) = \sum_{n=0}^\infty {1\over n!} \sum_{I_1,\cdots ,I_n\in \I_0^q}
{\partial^n f\over \partial \x^{I_1}\cdots
  \partial \x^{I_n}} (x,0) \x^{I_1}\cdots \x^{I_n}.
\]
Now $f\in {\cal I}^q_{pol}(|\Omega|)$ if and only if, $\forall I\in \I^q_2$,
\[
\forall x\in |\Omega|, \quad
{\partial f\over \partial \x^I}(x,0) = - \sum_{n=2}^\infty {1\over n!}
\sum_{I_1,\cdots ,I_n\in \I_0^q} \epsilon^{I_1\cdots I_n}_I
{\partial^n f\over \partial \x^{I_1}\cdots \partial \x^{I_n}} (x,0).
\]
Hence for such a function
\[
f(x,\x) = \sum_{n=2}^\infty {1\over n!} \sum_{I_1,\cdots ,I_n\in \I_0^q}
{\partial^n f\over \partial \x^{I_1}\cdots \partial \x^{I_n}}(x,0) \left[
  \x^{I_1}\cdots \x^{I_n} - \sum_{I\in \I^q_0} \epsilon^{I_1\cdots
    I_n}_I \x^I\right].
\]
So ${\cal I}^q_{pol}(|\Omega|)$ is the ideal spanned by the family
\[
\left( \x^{I_1}\cdots \x^{I_n} - \sum_{I\in \I^q_0} \epsilon^{I_1\cdots
    I_n}_I \x^I \right)_{n\geq 2,\  I_1,\cdots ,I_n\in \I^q_0}.
\]
Hence it is clear that the linear application from $\hbox{Span}_{{\cal
    C}^\infty(|\Omega|)}(\x^I)$ to $\hbox{Span}_{{\cal
    C}^\infty(|\Omega|)}(\eta^I)$ which maps $\x^I$ to $\eta^I$ 
can be extended in an unique way into
an algebra {\em
    isomorphism} from ${\cal A}^q(|\Omega|)$ to ${\cal C}^\infty(|\Omega|)\otimes \R[\eta^1,\cdots
,\eta^q]_0$. Moreover this isomorphism is nothing but
\[
\begin{array}{cccc}
\iota^* \circ e^{\vartheta} : & {\cal A}^q(|\Omega|) &
\longrightarrow & {\cal C}^\infty(|\Omega|)\otimes \R[\eta^1,\cdots
,\eta^q]_0 \\
& f & \longmapsto &  \iota^* \circ \left(e^{\vartheta}f\right)
\end{array}
\]

\subsection{An alternative description using schemes}
Let us start by assuming that $p=0$ for simpliclity.
The "geometry" of $\R^{0|q}$ appears to be related with another
"geometric" object living in a neighbourhood of 0 in
$\Lambda^{2*}_+\R^q$ and such that the ring of functions on it is
isomorphic to the algebra ${\cal A}^q:= {\cal A}^q(\{0\})$ that we just constructed. It
turns out that this object can be described accurately by using Grothendieck's
theory of schemes. We refer to \cite{EisenbudHarris} for a complete
and comprehensive presentation of this theory and recall here only
notions which may be relevant for us. To any commutative ring $R$ we can associate
an (affine) scheme which is called the {\em spectrum} of $R$ and is denoted
by $\hbox{Spec}R$. It consists in three data: a set
of points, a topology (the Zariski topology) and a sheaf of regular
functions on it. The set of points is
simply the set of prime ideals of $R$. In the case at hand where $R =
{\cal A}^q$ the prime ideals are of the form\footnote{here if $a_1,\cdots ,a_p\in R$,
  we denote by $(a_1,\cdots ,a_p)$ the ideal $\{a_1f_1 
+\cdots + a_pf_p| f_1,\cdots ,f_p\in R\}$}
\[
\mathfrak{A} = 
\left(\sum_{I\in \I^q_2}\alpha_{1,I}\x^I,\cdots ,\sum_{I\in
  \I^q_2}\alpha_{p,I}\x^I\right),
\]
where $p\in \N$ and the $\alpha_{j,I}$ are real parameters so that,
$\forall f,g\in R$, if $fg\in \mathfrak{A}$ then either $f\in
\mathfrak{A}$ or $g\in \mathfrak{A}$.
The "point" which
corresponds to such an ideal is the "generic point" living in the
vector subspace defined by $\sum_{I\in \I^q_2}\alpha_{1,I}\x^I = \cdots
=\sum_{I\in \I^q_2}\alpha_{p,I}\x^I = 0$. Note that by dualizing the canonical
ring morphism ${\cal C}^\infty_{pol}(\Lambda^{2*}_+\R^q)\longrightarrow {\cal
  A}^q$ we can view $\hbox{Spec}{\cal A}^q$ as embedded in
$\Lambda^{2*}_+\R^q$.
\begin{exem}\label{exampleofpoints}
For all $q\in \N$, set ${\cal A}_{(2)}^q:= \{\sum_{1\leq i\leq j<q}
\alpha_{ij}\x^{ij}\}$. Then for any $1\leq p\leq {q(q-1)\over 2}$ if
$f_1,\cdots ,f_p$ are $p$ linearly independants vectors of
${\cal A}_{(2)}^q$, then $(f_1,\cdots ,f_p)$ is a prime ideal of ${\cal
  A}^q$ (and for $p = {q(q-1)\over 2}$ it is the maximal ideal, see below). For
$q\leq 4$ there are no other prime ideals. However for $q\geq 5$ other
instances of prime ideal exist like $(\x^{1234} + \x^{15})$ for $q = 5$.
\end{exem}
So in general the concept of a "point" of a
scheme is different from the usual one, except if the point
is a maximal ideal. For $R ={\cal A}^q$ there is only one
maximal ideal\footnote{rings with an unique maximal ideal are {\em called
  local rings}} which is $(\x^I)_{I\in \I^q(2)}$: it corresponds to the
point $0\in \Lambda^{2*}_+\R^q$. This point is also the unique closed
point for the Zariski topology, all the other ones are open\footnote{Then
$R$ can be interpreted as the ring of functions on the points of
$\hbox{Spec}R$: to each prime ideal $\mathfrak{A}$ of $R$ we
associate the {\em residue field} $R/\mathfrak{A}$ and each $f\in R$ has
an image $[f \hbox{ mod }\mathfrak{A}$] in $R/\mathfrak{A}$ through the
canonical projection, so each $f\in R$ is identified with the "map"
\[
\begin{array}{cccc}
f: &\hbox{Spec}R & \longrightarrow & \hbox{residue fields}\\
 & \mathfrak{A} & \longmapsto & [f \hbox{ mod }\mathfrak{A}].
\end{array}
\]
Here we can interpret $[f \hbox{ mod }\mathfrak{A}]$ as being
isomorphic to the set of functions on the zero set of
all functions contained in $\mathfrak{A}$.
A more refined description of functions on $\hbox{Spec}R$ is given by the construction of a sheaf
${\cal O}_{\footnotesize \hbox{Spec}R}$ on the topological
space $\hbox{Spec}R$ such that the ring of global sections of ${\cal
  O}_{\footnotesize \hbox{Spec}R}$ is $R$ (see
\cite{EisenbudHarris}).}.\\


\noindent For $p\geq 1$, similarly we can associate to any open subset
$\Omega$ of $\R^{p|q}$ the scheme associated with ${\cal
  A}^q(|\Omega|)$, and we can picture its spectrum $\hbox{Spec}{\cal
  A}^q(|\Omega|)$ as an object embedded in $|\Omega|\times \Lambda^{2*}_+\R^q$. Then
we can interpret our results as follows: first for any
morphism $\phi^*:{\cal C}^\infty({\cal N}) \longrightarrow
{\cal C}^\infty(\Omega)_0$ we have found that there exists a family of maps
$\Phi:|\Omega|\times\Lambda^{2*}_+\R^q \longrightarrow {\cal N}$ (a class of maps modulo
$\sim$) such that $\Phi^*_{|\circ} = \phi^*$. We can simply denote by
$\Phi_{|\circ} = \phi$ this relation. Second through the algebra
isomorphism $\iota^*\circ e^\vartheta:{\cal
  A}^q(|\Omega|)\longrightarrow {\cal C}^\infty(\Omega)_0$
constructed in the previous section, we can
decompose $\Phi^*_{|\circ} = \iota^*\circ e^\vartheta \circ \Phi^*_{|\star}$, where
\[
\left(\Phi^*_{|\star}f\right)(x,\x):= \sum_{I\in \I_0^q}\x^I {\cal D}_I(f\circ \Phi)(x,0) =
\left(e^{\sum_{I\in \I^q_2}\x^I{\partial \over \partial \x^I}} f\circ
\Phi \right)(x,0) = (f\circ \Phi)(x,\x) \quad \hbox{mod }{\cal I}^q(|\Omega|).
\]
Hence $\Phi_{|\star}$ can be thought as a restriction of $\Phi$ to
$\hbox{Spec}{\cal A}^q(|\Omega|)$. Moreover if we denote by $T_\Omega$
the isomorphism from $\hbox{Spec}{\cal C}^\infty(\Omega)_0$
to $\hbox{Spec}{\cal A}^q(|\Omega|)$ which is dual of $\iota^*\circ e^\vartheta$ we
can dualize the relation $\Phi^*_{|\circ} = \iota^*\circ e^\vartheta
\circ \Phi^*_{|\star}$ as $\phi = \Phi_{\circ} = \Phi_{|\star}\circ
T_\Omega$. All that can be summarized in the following
diagrams:\\
\hskip1cm
\xymatrix{\Omega \ar@{->}[d]_{T_\Omega} \ar@{->}[dr]^{\phi = \Phi_{|\circ}} &
  \\
\hbox{Spec}{\cal A}^q(|\Omega|) \ar@{->}[r]_{\Phi_{|\star}} & {\cal
  N}}
\hskip1cm
\xymatrix{{\cal C}^\infty(\Omega)_0 \ar@{<-}[d]_{\iota^*\circ
    e^\vartheta} \ar@{<-}[dr]^{\phi^*=\Phi ^*_{|\circ}} \\
{\cal A}^q(|\Omega|) \ar@{<-}[r]_{\Phi^*_{|\star}} & {\cal C}^\infty({\cal N}) }

\section{Supermanifolds}
The previous and provisional definition of $\R^{p|q}$ can be recast in
the more sophisticated language of ringed
space, then functions on such superspaces can be seen as sections of sheaves of
superalgebras. Let us recall the definition of a supermanifold according to
\cite{Leites}, \cite{Manin}, \cite{DeligneMorgan}, \cite{Vara}. First one defines the
space $\R^{p|q}$ to be the topological space $\R^p$ endowed with the
sheaf of real superalgebras ${\cal O}_{\R^{p|q}}$ whose sections are
smooth functions on open subsets of $\R^p$, with values in
$\R[\theta^1,\cdots ,\theta^q]$, where $\theta^1,\cdots
,\theta^q$ are odd variables. So for any open subset $|\Omega|$ of $\R^p$ the
superalgebra $\Gamma(|\Omega|,{\cal O}_\Omega)$ of sections of ${\cal
  O}_{\R^{p|q}}$ over $|\Omega|$ is spanned over ${\cal C}^\infty(\Omega)$ by
$\theta^1,\cdots ,\theta^q$: $\forall f\in \Gamma(|\Omega|,{\cal
  O}_\Omega)$, $f = \sum_{I\in \I^q}f_I\theta^I$, where $f_I\in
{\cal  C}^\infty(|\Omega|)$, $\forall I\in\I^q$. The open subsets of ${\cal M}$
are then the objects $\Omega = (|\Omega|, {\cal O}_\Omega)$, where $|\Omega|$ is an
open subset of $\R^p$. If $\Omega$ and ${\Omega'}$ are two such open subsets then a
{\em morphism} $\varphi:\Omega \longrightarrow {\Omega'}$ is given by a
continuous map $|\varphi| : |\Omega| \longrightarrow |{\Omega'}|$ and
an even morphism $\varphi^*$ of sheaves of superalbegras from $|\varphi|^*{\cal O}_{\Omega'}$ to ${\cal
  O}_\Omega$\footnote{then, when restricted to the subsheaf ${\cal
    O}_{|{\Omega'}|}$ of smooth 
functions on $|{\Omega'}|$, $\varphi^*$ it corresponds to the usual pull-back operation on
functions by $|\varphi|$} (this implies in particular that $|\varphi|$ should
be smooth). If furthermore $|\varphi|$ is a {\em homeomorphism} and
$\varphi^*$ is an isomorphism of sheaves we then say that $\varphi$ is
an {\em isomorphism}.\\

\noindent
A supermanifold ${\cal M}$ of dimension $p|q$ is a topological space
$|{\cal M}|$ endowed with a sheaf ${\cal O}_{\cal M}$ of real
superalgebras which is {\em locally isomorphic} to $\R^{p|q}$. An {\em open
subset} $U$ of ${\cal M}$ is an open subset $|U|$ of $|{\cal
  M}|$ endowed with the sheaf of superalgebras ${\cal O}_U$ which
is the restriction of ${\cal O}_{\cal M}$  over $|U|$. By saying
{\em locally isomorphic} we mean that for any point $m\in |{\cal M}|$
there is an open subset $U$ of ${\cal M}$ such that $m\in |U|$, an
open subset $V$ of $\R^{p|q}$ and a isomorphism of sheaves $X$
from $U$ to $V$. There is however a difference with 
$\R^{p|q}$: the sheaf ${\cal O}_{|U|}$ of smooth real valued functions on
$|U|$ is not embedded in a canonical way in ${\cal
  O}_U$\footnote{i.e. by dualizing there is no canonical fibration ${\cal
  M}\longrightarrow |{\cal M}|$}. But it may be identified with ${\cal O}_U/{\cal
  J}$, where ${\cal J}$ is the nilpotent ideal $(\theta^1,\cdots
,\theta^q)$\footnote{i.e. by dualizing the projection map ${\cal O}_{\cal M}
\longrightarrow {\cal O}_{\cal M}/{\cal J}$, there is a canonical
embbeding $|{\cal M}|\hookrightarrow {\cal M}$}. Then the isomorphism
$X:U\longrightarrow V$ plays
the role of a local chart and the pull-back image of the canonical
coordinates $x^1,\cdots ,x^p,\theta^1,\cdots ,\theta^q$ by $X$
are the analogues of local coordinates.

\subsection{Maps from an open subset of $\R^{p|q}$ to a supermanifold}
Let ${\cal N}$ be a supermanifold of dimension
$n|m$, $U$ be an open subset of ${\cal N}$ and $Y:U\longrightarrow
V\subset \R^{n|m}$ be a local chart (i.e. a sheaf isomorphism). Let
$y^1,\cdots ,y^n,\psi^1,\cdots ,\psi^m$ be the canonical coordinates
on $\R^{n|m}$. By abusing notations we write also $y^\alpha:\simeq
Y^*y^\alpha$ and $\psi^j:\simeq Y^*\psi^j$. Then any section
$f$ of ${\cal O}_{\cal N}$ over $U$ decomposes as
\[
f = \sum_{J\in \I^m_0}F_J(y^1,\cdots ,y^n)\psi^J,
\]
where $\forall J\in \I^m_0$, $F_J\in {\cal C}^\infty(|V|)$ and
$\forall J = (j_1,\cdots ,j_k)$, $\psi^J:= \psi^{j_1}\cdots \psi^{j_k}$.\\

\noindent
Now let $\Omega$ be an open subset of $\R^{p|q}$ and $\phi$ be a map
from $\Omega$ to $U$, i.e. by dualizing an even morphism 
$\phi^*$ of superalgebra from ${\cal C}^\infty(U)$ to ${\cal
  C}^\infty(\Omega)$. Then the morphism property of $\phi^*$ implies
that
\[
\phi^*f = \sum_{J\in \I^m_0}\phi^*\left(F_J\circ (y^1,\cdots ,y^n)\right)
\chi^J,
\]
where $\forall j\in [\![1,m]\!]$, $\chi^j:= \phi^*\psi^j$, $\forall
J = (j_1,\cdots ,j_k)$, $\chi^J:= \chi^{j_1}\cdots \chi^{j_k}$ and
each $\phi^*\left(F_J\circ (y^1,\cdots ,y^n)\right)$ can be expressed
in terms of $(\phi^*y^1,\cdots , \phi^*y^n)$ by using Proposition
\ref{proptaylor}. Hence {\em $\phi^*f$ can be computed as soon as we
  know $(\phi^*y^1,\cdots , \phi^*y^n)$ and $(\phi^*\psi^1,\cdots ,
  \phi^*\psi^m)$}. This generalizes Proposition \ref{proptaylor}.

\subsection{The use of the functor of point}
When we study supersymmetric differential equations, a brutal
application of the previous definitions suffers from
incoherences. These are largely discussed in \cite{Freed}. An instance is the
superspace formulation of supergeodesics on an Euclidean sphere
$S^n$. Let us view $S^n$ as a submanifold of $\R^{n+1}$ and we consider
the "supertime" $\R^{1|1}$ with coordinates $t,\theta$. Then we look
at maps $\phi:\R^{1|1}\longrightarrow S^n$ (i.e. morphisms
$\phi^*$ from ${\cal C}^\infty(S^n)$ to ${\cal C}^\infty(\R^{1|1})$) which are solutions of
\[
D{\partial \phi\over \partial t} + \left\langle D\phi,{\partial
    \phi\over \partial t}\right\rangle \phi = 0,
\]
where $D:= {\partial \over \partial \theta} - \theta{\partial \over
  \partial t}$. This means that
the image of any coordinate function $y^\alpha$ on $\R^{n+1}\supset
S^n$ by $D{\partial \phi^*\over \partial t} + \left\langle D\phi^*,{\partial
    \phi^*\over \partial t}\right\rangle \phi^*$ vanishes. Set
  $\phi^*y = \varphi + \theta \psi$, where $\varphi\in {\cal
  C}^\infty(\R,S^n)$ and $\psi$ is a section of $\varphi^*TS^n$. A
  first problem is that $\psi$ should be {\em odd}: this is the usual
  requirement made by physicists and in our context it is imposed by
  the fact that $\phi^*$ should be an even morphism, because $\theta$
  is odd. This could be cared by introducing a further (dumb) odd
  variable, say $\eta$, and by letting $\psi = \eta v$, where $v$ is
  an ordinary section of $\varphi^*TS^n$. But then the next problem is
  that the preceding equation is equivalent to the
  system
\[
{\partial^2 \varphi\over (\partial t)^2} + \left|{\partial \varphi\over
    \partial t}\right|^2 \varphi = - \left\langle\psi, {\partial \varphi\over
    \partial t}\right\rangle \psi
\quad\hbox{and}\quad
{\partial \psi\over \partial t} + \left\langle\psi, {\partial \varphi\over
    \partial t}\right\rangle \varphi = 0.
\]
And we see that the right hand side of the first equation contains two times
$\psi$, hence $\eta\eta$, which vanishes. So we should build $\psi$ out of a linear
combination of at least two dumb odd variables, say $\eta^1$ and
$\eta^2$. But then we see that $\varphi$ cannot be an ordinary map
into $S^n$, still because of the first equation. Note that all these difficulties are
absent in the differential geometric point of view used in
\cite{Dewitt,Rogers} for defining supermanifolds.\\

\noindent An alternative solution is proposed in \cite{DeligneMorgan} and \cite{FP2}
(see also \cite{Vara}): it relies on Grothendieck's notion of {\em functor of points} in
algebraic geometry. We will adopt that point of view in the
following. For any $L\in \N$ we set $B:= \R^{0|L}$. The 
starting point is to see a map $\phi$ from a supermanifold ${\cal M}$
of dimension $p|q$ into a supermanifold ${\cal N}$ of dimension $n|m$
as {\em a functor from ${\cal C}^\infty(B)$ to even morphisms $\phi^*:{\cal C}^\infty({\cal N})
\longrightarrow {\cal C}^\infty({\cal M}\times B)$}. So we need to
understand morphisms $\phi^*$ from ${\cal C}^\infty({\cal N})$
to ${\cal C}^\infty({\cal M}\times B)$: from a technical point of view
nothing is new and it suffices to apply all the previous results. For simplicity we restrict
ourself to the case where the target manifold ${\cal N}$ is an
ordinary manifold and the source domain $\Omega$ is an open subset of
$\R^{p|q}$.

\subsection{Our final representation of a map from an open subset of
  $\R^{p|q}$ to an ordinary manifold}
It is convenient to note $(x^1,\cdots ,x^p)$, $(\theta^1,\cdots , \theta^q)$
respectively the even and the odd local coordinates on $\Omega$ and
$(\eta^1,\cdots , \eta^L)$ the odd coordinates on $B$. Hence for any
open subset $\Omega$ of ${\cal M}$, ${\cal
  C}^\infty(\Omega\times B) \simeq {\cal C}^\infty(|\Omega|)[\theta^1,
\cdots ,\theta^q, \eta^1,\cdots ,\eta^L]$. Furthermore we note $\A^q(0) =
\{\emptyset \}$ and for any $k\in \N^*$,
$\A^q(k):= \{(a_1,\cdots a_k)\in [\![1,q]\!]^k| a_1<\cdots <a_k\}$. We
denote by $A=(a_1,\cdots a_{k})$ an element of 
$\A^q(k)$ and we then write $\theta^A:= \theta^{a_1}\cdots
\theta^{a_k}$. And we let $\A^q:= \cup_{k=0}^q\A^q(k)$, $\A^q_0:=
\cup_{k=0}^{[q/2]}\A^q(2k)$, $\A^q_1:= \cup_{k=0}^{[(q-1)/2]}\A^q(2k+1)$,
$\A^q_2:= \cup_{k=1}^{[q/2]}\A^q(2k)$ and $\A^q_+:= \A^q_1\cup \A^q_2$. Lastly we set $\AI:=
\{AI|A\in \A^q,I\in \I^L\}$ and, defining the degree of $AI$ to be the
some of the degrees of $A$ and $I$, we define similarly $\AI(j)$, $\AI_0$,
$\AI_1$ and $\AI_2$. Hence any (even) function $f\in
{\cal C}^\infty(\Omega\times B)$ (where $\Omega$ is an open subset of
$\R^{p|q}$) can be decomposed as $f = \sum_{AI\in \AI_0}
\theta^A\eta^If_{AI},$
where $f_{AI}\in {\cal C}^\infty(|\Omega|)$, $\forall AI\in
\AI_0$.\\

\noindent
Then Theorem \ref{letheo} implies that for any morphism
$\phi^*$ from ${\cal C}^\infty({\cal N})$ to ${\cal
  C}^\infty(\Omega\times B)$, there exists a smooth map
$\varphi\in {\cal C}^\infty(|\Omega|,{\cal N})$ and  a smooth family
$\left(\xi_{AI}\right)_{AI\in \AI_2}$ of sections of $\pi^*T{\cal N}$
defined on a neighbourhood of the graph of $\varphi$ in
$|\Omega|\times {\cal N}$ such that if $\Xi:= \sum_{AI\in
  \AI_2}\xi_{AI}\theta^A\eta^I$ then $\forall f\in {\cal
  C}^\infty({\cal N})$, $\phi^*f = (1\times f)^*\left(e^\Xi
  f\right)$. Moreover, thanks to Proposition \ref{prop1}, the vector
fields $\left(\xi_{AI}\right)_{AI\in \AI_2}$ can be chosen in order to
commute pairwise. We decompose $\Xi$ as
\[
\Xi = \sum_{A\in \A^q}\theta^A\Xi_A = \Xi_\emptyset + \sum_{a\in
  \A^q(1)} \theta^a\Xi_a + \sum_{(a_1,a_2)\in \A^q(2)}
\theta^{a_1} \theta^{a_2}\Xi_{a_1a_2} + \cdots ,
\]
where $\forall A\in \A^q_1$, $\Xi_A = \sum_{I\in
  \I^L_1}\xi_{AI}\eta^I$ and $\forall A\in \A^q_0$, $\Xi_A = \sum_{I\in
  \I^L_2}\xi_{AI}\eta^I$. In particular $\Xi_\emptyset = \sum_{I\in
  \I^L_2}\xi_{\emptyset I}\eta^I$ and we see that $\Xi_A$ is odd if $A$ is odd
  and is even if $A$ is even. Then the relations
  $[\xi_{AI},\xi_{A'I'}] = 0$ implies that the
vector fields $\Xi_A$ {\em supercommute} pairwise, i.e.
\[
\forall A\in \A^q(k),\forall A'\in \A^q(k'),\quad
\Xi_A\Xi_{A'} - (-1)^{kk'} \Xi_{A'}\Xi_A = 0.
\]
This is equivalent to the
fact that $\forall A,A'\in \A^q$,  $[\theta^A\Xi_A,
\theta^{A'}\Xi_{A'}] = 0$. This last commutation relation implies that
\[
e^\Xi = e^{\sum_{A\in \A^q}\theta^A\Xi_A} = e^{\Xi_\emptyset}
\prod_{A\in \A^q_+} e^{\theta^A\Xi_A} = e^{\Xi_\emptyset}
\prod_{A\in \A^q_+} (1 + \theta^A\Xi_A),
\]
where we have used $\left(\theta^A\Xi_A\right)^2 = 0$. Hence
\begin{equation}\label{finalenxi}
\forall f\in{\cal C}^\infty({\cal N}),\quad
\phi^*f = (1\times \varphi)^*\left(e^{\Xi_\emptyset}
\prod_{A\in \A^q_+} (1 + \theta^A\Xi_A)f\right).
\end{equation}
Alternatively one can integrate these vector fields as in the second
section of this paper. Let us denote by $\left(
  \x^{AI}\right)_{AI\in \AI_2}$ the coordinates on $\Lambda^{2*}_+\R^{q+L}$
and
\[
\vartheta:= \sum_{AI\in \AI_2}\theta^A\eta^I{\partial \over \partial
  \x^{AI}} = \sum_{A\in \A^q} \theta^A\vartheta_A,
\]
where $\forall A\in \A^q_1$, $\vartheta_A:= \sum_{I\in
  \I^L_1}\eta^I{\partial \over \partial \x^{AI}}$ and $\forall A\in \A^q_0$, $\Xi_A = \sum_{I\in
  \I^L_2}\eta^I{\partial \over \partial \x^{AI}}$. Then there exists a
  smooth map $\Phi$ from a neighbourhood of $|\Omega|\times\{0\}$ in
  $|\Omega|\times\Lambda^{2*}_+\R^{q+L}$ to ${\cal N}$ such that
\begin{equation}\label{finalens}
\forall f\in{\cal C}^\infty({\cal N}),\quad
\phi^*f = \iota^*\left(e^{\vartheta_\emptyset}
\prod_{A\in \A^q_+} (1 + \theta^A\vartheta_A)(f\circ \Phi)\right).
\end{equation}

\subsection{Forgetting the ugly notations}
We now propose some abuses and adaptations of notation to lighten all this
description. But we try to keep the important property
that each $\Xi_A$ is vector field\footnote{i.e. a {\em first order}
  differential operator} defined along the graph of $\varphi$ (even if it
has coefficients in a Grassmann algebra). First of all we simply
write $\varphi^*:\simeq (1\times \varphi)^*$. Second the
operator $e^{\Xi_\emptyset}$ has no direct geometrical signification
and his presence there is only necessary to "thicken" $\varphi^*$, so
that we can absorb it by a redefinition of $\varphi^*$:
\[
\varphi^*:\simeq \varphi^*e^{\Xi_\emptyset} :\simeq (1\times \varphi)^* e^{\Xi_\emptyset}.
\]
We can hence rewrite (\ref{finalenxi}) as

\begin{equation}\label{finalsimplexi}
\forall f\in {\cal C}^\infty({\cal N}),\quad
\phi^*f = \varphi^*\left( \prod_{A\in \A^q_+}(1 +
  \theta^A\Xi_A)\right) f.
\end{equation}
For example if $q = 2$, we have (keeping in mind the fact that $\Xi_1$
and $\Xi_2$ are odd whereas $\Xi_{12}$ is even):
\begin{equation}\label{exam0}
\begin{array}{ccl}
\phi^*f & = & \varphi^*
(1+\theta^1\Xi_1)(1+\theta^2\Xi_2)(1+\theta^1\theta^2\Xi_{12})f\\
& = &  \varphi^* \left(1
+ \theta^1\Xi_1+\theta^2\Xi_2+\theta^1\theta^2(\Xi_{12} -
\Xi_1\Xi_2)\right)f.
\end{array}
\end{equation}
Similarly relation (\ref{finalens}) can be written
\begin{equation}\label{finalsimples}
\forall f\in {\cal C}^\infty({\cal N}),\quad
\phi^*f = \left( \prod_{A\in \A^q_+}(1 +
  \theta^A\vartheta_A)\right) \Phi^* f.
\end{equation}

\noindent
{\bf Use of a local chart on the target manifold}\\

\noindent
The use of relations (\ref{finalsimplexi}) is particularly convenient
if we assume that the image of $\phi:\Omega\longrightarrow {\cal N}$
is contained in an open subset $U\subset {\cal N}$ on which there is a
chart $y:U\longrightarrow \R^n$. Indeed Remark \ref{remaxixiy=0} tells us
that we can choose the vector fields $\left(\xi_I\right)$ in such a
way that $\xi_I\xi_Jy = 0$ (see (\ref{xixiy=0})). This implies that
$\Xi_A\Xi_{A'}y = 0$, $\forall A,A'\in \A^q$. Now what physicists
denote "$\phi$" or "$\left(\phi^\alpha\right)_\alpha$" is just $\phi^*y$ or
$\left(\phi^*y^\alpha\right)_\alpha$ and then when they write the decomposition
\begin{equation}\label{fromphysics}
"\phi = \varphi + \sum_{A\in \A^q}\theta^A\psi_A",
\end{equation}
it implies by using (\ref{finalsimplexi}) that
\[
\varphi + \sum_{A\in \A^q_+}\theta^A\psi_A = \phi^*y
= \varphi^*\left( \prod_{A\in \A^q_+}(1 +
  \theta^A\Xi_A)\right)y.
\]
But since $\Xi_A\Xi_{A'}y = 0$ the development of the right hand side of
this identity is particularly simple. We deduce
\[
\varphi + \sum_{A\in \A^q_+}\theta^A\psi_A = 
\varphi^* y + \sum_{A\in \A^q_+}\theta^A\varphi^*\Xi_Ay.
\]
Hence $\forall A\in \A^q_+$, $\psi_A = \varphi^*\Xi_Ay$. Our last abus
of notation is to let $\psi_A\simeq \Xi_A$. So we reinterpret
(\ref{fromphysics}) as
\[
\phi^* = \varphi^* \prod_{A\in \A^q_+} (1 + \theta^A\psi_A),
\]
where the rules to manipulate such an expression are
\begin{itemize}
\item each $\psi_A$ acts as a first order differential operator to its
  right
\item two different $\psi_A$, $\psi_{A'}$ supercommute pairwise and
  with the $\theta^A$'s
\end{itemize}

\noindent
{\bf An example of application}\\

\noindent
Assume that we find in the physics litterature  a map "$\phi$" from $\R^{p|2}$
to $\R$ which has the expression
\begin{equation}\label{exam1}
"\phi = \varphi + \theta^1\psi_1 +
\theta^2\psi_2 +\theta^1\theta^2 F"
\end{equation}
and we want to compute $\phi^*f\simeq f\circ \phi$, where $f\in {\cal
  C}^\infty(\R)$. Then we reinterpret (\ref{exam1}) as
\[
\phi^* = \varphi^*(1+\theta^1\psi_1) ( 1 + \theta^2\psi_2) (1 +
\theta^1\theta^2F).
\]
Then
\[
\begin{array}{ccl}
\phi^*f & = & \varphi^*(1+\theta^1\psi_1) ( 1 + \theta^2\psi_2) (1 +
\theta^1\theta^2F)f \\
 & = & \varphi^* f + \theta^1 \varphi^*\psi_1f + \theta^2 \varphi^*\psi_2f +
 \theta^1\theta^2\varphi^* Ff - \theta^1\theta^2\varphi^*\psi_1\psi_2f \\
 & = & f\circ \varphi + \theta^1 (f'\circ \varphi)\psi_1 + \theta^2
 (f'\circ \varphi)\psi_2 +  \theta^1\theta^2 [(f'\circ \varphi)F -
 (f''\circ \varphi) \psi_1\psi_2].
\end{array}
\]

\end{document}